\documentclass[12pt,onecolumn,amsmath,amssymb]{revtex4}

\begin{document}

\title{Dynamics of the Free Surface of a Conducting Liquid
in a Near-Critical Electric Field}

\author{Nikolay M. Zubarev}
\email{nick@ami.uran.ru}
\author{Olga V. Zubareva}

\affiliation{Institute of Electrophysics, Ural Branch, Russian
Academy of Sciences,\\ 106 Amundsen Street, 620016 Ekaterinburg, Russia}

\begin{abstract}

Near-critical behavior of the free surface of an ideally conducting
liquid in an external electric field is considered. Based on an analysis
of three-wave processes using the method of integral estimations, sufficient
criteria for hard instability of a planar surface are formulated. It is
shown that the higher-order nonlinearities do not saturate the
instability, for which reason the growth of disturbances has an explosive
character.
\end{abstract}

\maketitle

\section{INTRODUCTION}

The electrohydrodynamic instability of the free surface
of a conducting liquid in a strong electric field
\cite{1,2} is responsible for many physical processes, such
as the initiation and maintenance of emission of
charged particles, vacuum breakdown, vacuum discharge,
etc. The interaction of an electric field and
charges induced by this field on the surface of a liquid
(liquid metal) leads to a growth of surface disturbances
and the formation of regions with a significant curvature
\cite{3,4,5}. The dispersion law for the waves on a planar
surface of an ideally conducting liquid in an external
electric field of strength $E$
has the following form \cite{6}:
\begin{equation}
\omega^2=g|{\bf k}|+\frac{\alpha}{\rho}\,|{\bf k}|^3-
\frac{E^2}{4\pi\rho}\,|{\bf k}|^2
\end{equation}
where $\omega$ is the frequency, ${\bf k}$ is the wave vector, $g$
is the acceleration of gravity, $\alpha$ is the surface tension
coefficient, and $\rho$ is the density of the medium.

It is seen from Eq.~(1) that if the condition
$$
E^2<{E_c}^2=8\pi\sqrt{g\alpha\rho},
$$
is fulfilled, then $\omega^2>0$ at any $|{\bf k}|$
and, consequently, the
surface disturbances do not grow with time.
If the magnitude of the field $E$, which plays the role of an external
governing parameter, exceeds a certain critical value $E_c$,
then there arises a region of wave vectors $|{\bf k}|$,
for which $\omega^2<0$, which corresponds to an aperiodic instability.
Thus, the condition $E>E_c$ is a criterion for the
surface instability with respect to infinitely small disturbances
of the surface shape and of the field of velocities.

It was shown in \cite{7,8,9}, where liquids with various
physical properties have been considered, that a nonlinear
interaction of three standing waves that form a hexagonal
structure can lead to a hard excitation of the
instability of a charged surface. In our case, this means
that, even at subcritical fields ($E<E_c$)
a disturbance of
a sufficient magnitude can break the equilibrium of a
planar surface. In this connection, there arises a need of
constructing criteria for the instability of a charged surface
of a conducting liquid with respect to perturbations
of a finite magnitude, i.e., criteria that will permit one,
proceeding from some initial data such as the shape of
the surface and the distribution of velocities, to answer
the question of whether or not the initial perturbation
will lead to the loss of the stability of a planar boundary
and, as a consequence, to an explosive growth of cusplike
structures. This work is devoted to constructing
such criteria using the method of integral estimations
that was applied previously to obtain the conditions of
collapse for the nonlinear Schrodinger equation
\cite{10,11}, nonlinear Klein-Gordon equation \cite{12,13},
and various modifications of the Boussinesq equation
\cite{14}.

In Section 2, we give equations of the vertex-free
motion of an ideally conducting liquid with a free surface
in an electric field and give their Hamiltonian formulation.
In Section 3, the theory of perturbations in a
small parameter, namely, in the characteristic angle of
the surface slope, is constructed up to fourth-order
terms in the Hamiltonian. The analysis of the surface
dynamics is significantly simplified in the case of small
"supercriticalities" (i.e., if the field only slightly
exceeds the critical value $E_c$)
$$
\varepsilon=(E^2-E_c^2)/E_c^2, \qquad |\varepsilon|\ll 1,
$$
when only perturbations with wave numbers close to
$k_0=\sqrt{g\rho/\alpha}$ increase (this value of the wave number
corresponds to the so-called dominant harmonic of surface
perturbations).
This permits us in Section 4 to construct
a set of amplitude equations for describing the
nonlinear interaction of three standing waves that form
a hexagonal structure which is the main interaction at near-critical
values of the field $E$.
In Section 5, we
extend the method of integral estimations for several
interacting nonlinear waves. By using this method, one
can pass from a set of partial differential equations for
complex amplitudes $A_1$, $A_2$ and $A_3$ to
to a second-order differential inequality for the norm
$$
X=\int \left(|A_1|^2+|A_2|^2+|A_3|^2\right)d^2r,
$$
by analyzing which we obtain a number of sufficient
criteria of hard excitation of an electrohydrodynamic
instability of a charged surface. Note that most of them
refer to subcritical values of the external electric field,
when the surface is stable in a linear approximation,
while the development of the instability is related to
three-wave processes. In Section 6, we show, on the
example of one-dimensional and square lattices of surface
perturbations for which the three-wave interactions
degenerate, that the higher-order wave processes
do not saturate the instability but, on the contrary, lead
to an explosive growth of amplitudes.

\section{STARTING EQUATIONS}
Consider the potential motion of an ideal conducting
liquid of infinite depth placed in an external uniform
electric field of strength $E$.
Assume that the fieldstrength vector is directed along the $z$
axis and, correspondingly, in the unperturbed state the boundary of the
liquid is a planar horizontal surface $z=0$.
Let the function $\eta(x,y,t)$
specify the deviation of the boundary
from the flatness, i.e., the region occupied by the liquid
is restricted by a free surface $z=\eta$.

The velocity potential for an incompressible liquid $\Phi$
satisfies Laplace's equation
\begin{equation}
\nabla^2\Phi=0,
\end{equation}
with the following conditions at the metal-vacuum boundary and at infinity:
$$
\frac{\partial\Phi}{\partial t}+\frac{(\nabla\Phi)^2}{2}=
\frac{(\nabla\varphi)^2-E^2}{8\pi\rho}+
\frac{\alpha}{\rho}\,\nabla_{\!\!\bot}\cdot
\frac{\nabla_{\!\!\bot}\eta}{\sqrt{1+(\nabla_{\!\!\bot}\eta)^2}}-
g\eta, \qquad z=\eta,
$$
\begin{equation}
\Phi\to 0, \qquad z\to-\infty,
\end{equation}
where $\varphi$ is the electric-field potential.

The first term on the right-hand side of the dynamic
boundary condition (nonstationary Bernoulli equation)
is responsible for the electrostatic pressure, the second
term determines the capillary pressure, and the third
term takes into account the effect of the gravitational
field. The time evolution of the free surface is determined
by the kinematic relationship (condition of non-
flowing of the liquid through its boundary)
$$
\frac{\partial\eta}{\partial t}=\frac{\partial\Phi}{\partial
z} -\nabla_{\!\!\bot}\eta\cdot\nabla_{\!\!\bot}\Phi,
\qquad z=\eta.
$$
Finally, the electric-field potential $\varphi$
in the absence of spatial charges satisfies Laplace's equation
$$
\nabla^2\varphi=0,
$$
which should be solved simultaneously with the condition
of the equipotentiality of the boundary of the conducting
liquid and the condition of the uniformity of the
field at an infinite distance from the surface:
$$
\varphi=0, \qquad z=\eta,
$$
$$
\varphi\to -Ez, \qquad z\to\infty.
$$

Note that the above-written equations of motion
have a Hamiltonian structure and the functions
$\eta(x,y,t)$ and $\psi(x,y,t)=\Phi|_{z=\eta}$
are canonically conjugated quantities \cite{15}
\begin{equation}
\frac{\partial\psi}{\partial t}=-\frac{\delta H}{\delta\eta},
\qquad
\frac{\partial\eta}{\partial t}=\frac{\delta H}{\delta\psi},
\end{equation}
where the Hamiltonian $H$
coincides to an accuracy of a
constant with the total energy of the system
$$
H=\int\limits_{z\leq\eta}\frac{(\nabla\Phi)^2}{2} d^3 r-
\int\limits_{z\geq\eta}\frac{(\nabla\varphi)^2}{8\pi\rho} d^3r
$$
$$
+\int\left[\frac{g\eta^2}{2}+
\frac{\alpha}{\rho}\left(\sqrt{1+(\nabla_{\!\!\bot}\eta)^2}-1
\right)\right]d^2 r.
$$

For a further consideration of the problem, it is convenient
to represent the Hamiltonian in the form of a
surface integral. We introduce a perturbation of the
electric-field potential $\tilde\varphi=\varphi+Ez$.
It can easily be shown that the perturbed potential $\tilde\varphi$
satisfies Laplace's equation
\begin{equation}
\nabla^2\tilde\varphi=0,
\end{equation}
with conditions
\begin{equation}
\tilde\varphi=E\eta, \qquad z=\eta,
\end{equation}
\begin{equation}
\tilde\varphi\to 0, \qquad z\to\infty,
\end{equation}
from which it is seen that the perturbation introduced
by the surface $z=\eta$ into the distribution of the electric
field decays as $z\to\infty$.
Taking into account that, in
view of the incompressibility of the liquid, a relation
$\int\tilde\varphi|_{z=\eta}d^2r=0$
is valid, and neglecting terms whose
variation does not contribute to the equation of motion,
we obtain, using the first Green's formula,
$$
H=\int\limits_{s}\left[\frac{\psi}{2}\,
\frac{\partial\Phi}{\partial n}+
\frac{E\eta}{8\pi\rho}\,
\frac{\partial\tilde\varphi}{\partial n}\right]ds+
\int\left[\frac{g\eta^2}{2}+
\frac{\alpha}{\rho}\left(\sqrt{1+(\nabla_{\!\!\bot}\eta)^2}-
1\right) \right]d^2r,
$$
where $ds$ is the surface differential and $\partial/\partial n$
denotes the derivative in the direction of the normal to the surface
$z=\eta$.

By eliminating the normal derivatives of the potentials
$\tilde\varphi$ and $\Phi$, we can reduce the expression for the
Hamiltonian to the form
$$
H=\int\limits\left[\frac{\psi}{2} \left.\left(\Phi_z-
\nabla\eta\cdot\nabla_{\!\!\bot}\Phi \right)\right|_{z=\eta}+
\frac{E\eta}{8\pi\rho}\left.\left(\tilde\varphi_z-
\nabla\eta\cdot\nabla_{\!\!\bot}\tilde\varphi\right)\right|_{z=
\eta} \right]d^2r
$$
\begin{equation}
+\int\left[\frac{g\eta^2}{2}+
\frac{\alpha}{\rho}\left(\sqrt{1+(\nabla_{\!\!\bot}\eta)^2}-
1\right)\right] d^2r,
\end{equation}
which is more suitable for further transformations.

\section{SMALL-ANGLE APPROXIMATION}

Our further problem is to eliminate the spatial variable $z$
from the equations of motion, i.e., to pass from
the initial three-dimensional equations to two-dimensional
ones. To do this, we should write the integrand in
Eq.~(8) through the canonical variables $\eta$ and $\psi$.
Then,
there arises a need to solve Eq.~(5) with conditions (6)
and (7) as well as Eq.~(2) with the condition
$$
\Phi=\psi, \qquad z=\eta,
$$
and condition (3). We use the known solutions to
Laplace's equation for the half-spaces
$z<0$ and $z>0$ for functions that decay at infinity
\begin{equation}
\tilde\varphi(x,y,z)=\frac{1}{2\pi}\!\int\limits_{-
\infty}^{+\infty} \int\limits_{-\infty}^{+\infty}
\frac{z\tilde\varphi(x',y',0)}{\left[(x'-x)^2+(y'-
y)^2+z^2\right]^{3/2}} \,dx'dy',\qquad z>0,
\end{equation}
\begin{equation} \Phi(x,y,z)=-\frac{1}{2\pi}\!\int\limits_{-
\infty}^{+\infty} \int\limits_{-\infty}^{+\infty}
\frac{z\Phi(x',y',0)}{\left[(x'-x)^2+(y'-y)^2+z^2\right]^{3/2}}
\,dx'dy',\qquad z<0.
\end{equation}
Now, we should express the magnitudes of the potentials
$\tilde\varphi$ and $\Phi$ that enter into these relationships at the
plane  $z=0$ through their magnitudes at the boundary $z=\eta$, i.e.,
through the functions $E\eta$ and $\psi$.
Let the characteristic angles of the surface slope be small:
$|\nabla_{\!\!\bot}\eta|\ll 1$.
In this case the potentials near the $z=0$ plane
can be expanded into a power series in surface perturbation
$\eta$:
\begin{equation}
\tilde\varphi(x,y,\eta(x,y))=\sum_{n=0}^{\infty}\frac{\eta^n
}{n!}\, \left.\frac{\partial^n\tilde\varphi}{\partial
z^n}\right|_{z=0}, \qquad
\Phi(x,y,\eta(x,y))=\sum_{n=0}^{\infty}\frac{\eta^n}{n!}\,
\left.\frac{\partial^n\Phi}{\partial z^n}\right|_{z=0}.
\end{equation}
By differentiating Eqs.~(9) and (10) with respect to $z$,
we find that
$$
\left.\tilde\varphi_z\right|_{z=0}=-
\hat{k}\tilde\varphi|_{z=0},
\qquad
\left.\Phi_z\right|_{z=0}=\hat{k}\Phi|_{z=0},
$$
where $\hat k$ is the two-dimensional integral operator given
by the expression
$$
\hat{k}f=-\frac{1}{2\pi}\!\int\limits_{-\infty}^{+\infty}
\int\limits_{-\infty}^{+\infty} \frac{f(x',y')}{\left[(x'-
x)^2+(y'-y)^2\right]^{3/2}}\,dx'dy'.
$$
This relationship can be considered as a consequence
of the fact that the Laplacian operator can formally
be represented as
$$ \nabla^2=\left(\partial_z+\hat{k}\right)\left(\partial_z-
\hat{k}\right), $$
where the left-hand bracket corresponds to solutions
that are asymptotically decay as
$z\to+\infty$ and the
right-hand bracket corresponds to solutions that decay
as $z\to-\infty$.

By eliminating the derivative with respect to $z$
from the expansions (11), we find
$$
\tilde\varphi(x,y,\eta(x,y))=\hat T_+\tilde\varphi(x,y,0),
\qquad \Phi(x,y,\eta(x,y))=\hat T_-\Phi(x,y,0),
$$
where we introduced nonlinear shear operators
$$
\hat T_+=\sum_{n=0}^{\infty}\frac{\eta^n\hat k^n}{n!},
\qquad
\hat T_-=\sum_{n=0}^{\infty}\frac{(-\eta)^n\hat k^n}{n!}.
$$
Let $T_\pm^{-1}$ be the operators that are inverse with
respect to the shear operators. Their form can be determined
using the method of successive approximations
$$
T_\pm^{-1}=1\mp\eta\hat{k}-
\eta^2\hat{k}^2/2+\eta\hat{k}\eta\hat{k}+...
$$
Then, we have
$$
\tilde\varphi(x,y,0)=E\hat T_+^{-1}\eta(x,y), \qquad
\Phi(x,y,0)=\hat T_-^{-1}\psi(x,y).
$$

These relationships, along with Eqs.~(9) and (10),
specify the solutions to Laplace's equations with necessary
boundary conditions in the form of infinite series.
Using these solutions, we can write the various possible
derivatives of the potentials $\tilde\varphi$ and $\Phi$
that enter into the Hamiltonian through the functions
$\eta$ and $\psi$. As a result, we find
$$
H=\int\frac{\psi}{2}\left(\hat T_+\hat k\hat T_+^{-1}\psi-
\nabla_{\!\!\bot}\eta\cdot\hat T_+\nabla_{\!\!\bot}
\hat T_+^{-1}\psi\right) d^2 r
$$
$$
-\int\frac{E^2\eta}{8\pi\rho}
\left(\hat T_-\hat k\hat T_-^{-1}\eta+
\nabla_{\!\!\bot}\eta\cdot\hat T_-\nabla_
{\!\!\bot}\hat T_-^{-1}\eta\right)d^2 r
$$
$$
+\int\left[\frac{g\eta^2}{2}+
\frac{\alpha}{\rho}\left(\sqrt{1+(\nabla_{\!\!\bot}\eta)^2}-
1\right)\right] d^2r.
$$

To describe the initial stages of instability development
on the surface of a conducting liquid, it is sufficient
to restrict ourselves by allowance for a finite number
of terms in the expansion of the integrands of the
functional $H$ in canonical variables. By omitting terms
of higher than the fourth order of smallness for the surface
perturbation $\eta$ and higher than the second order for
the potential $\psi$,
(this proves to be sufficient at small
supercriticalities) and successively integrating by parts,
we finally obtain
\begin{equation}
H=H^{(2)}+H^{(3)}+H^{(4)},
\end{equation}
$$
H^{(2)}=\int\left[\frac{\psi\hat{k}\psi}{2}-
\frac{E^2\eta\hat{k}\eta}{8\pi\rho}+
\frac{g\eta^2}{2}+\frac{\alpha(\nabla\eta)^2}{2\rho}
\right]d^2r,
$$
$$
H^{(3)}=\frac{E^2}{8\pi\rho}\int \eta\left[(\nabla\eta)^2-
(\hat{k}\eta)^2\right]d^2r,
$$
$$
H^{(4)}=-\frac{E^2}{8\pi\rho}
\int\left[\eta\hat{k}\eta\hat{k}\eta\hat{k}\eta+
\eta\hat{k}\eta^2\nabla^2\eta\right]d^2r -
\int\frac{\alpha(\nabla\eta)^4}{8\rho}\,d^2r.
$$
These expressions, in combination with Eq.~(4), represent
a two-dimensional reduction of the equations of
motion of a conducting liquid in an external electric
field that is applicable if the condition of the smallness
of the characteristic angles of the surface slope is fulfilled.

\section{AMPLITUDE EQUATIONS}

Now, let us consider the nonlinear dynamics of the
perturbations of the free surface of a conducting liquid
for the case where the magnitude of the external electric
field $E$ is close to its threshold value $E_c$, i.e.,
$|\varepsilon|\ll 1$.
It follows from the dispersion relation (1) that, at small
supercriticalities, only surface waves with wave numbers
close to $k_0$ can be excited. The main nonlinear
interaction in this case will be the three-wave interaction
between the waves whose wave vectors are turned
with respect to one another by an angle of $2\pi/3$.
This can easily be understood from the conditions
$$
{\bf k}_1+{\bf k}_2+{\bf k}_3=0, \qquad
|{\bf k}_1|=|{\bf k}_2|=|{\bf k}_3|=k_0.
$$
Near the threshold, it is natural to pass to envelopes
using the following substitutions:
$$
\eta({\bf r},t)=\sum_{j=1}^3
A_j (x_j,y_j,t)e^{i{\bf k}_j {\bf r}}+\mbox{c.c.},
$$
$$
\psi({\bf r},t)=\sum_{j=1}^3
B_j(x_j,y_j,t)e^{i{\bf k}_j {\bf r}}+\mbox{c.c.},
$$
where ${\bf k}_1=\{k_0,0\}$, ${\bf k}_2=\{-
k_0/2,\sqrt{3}k_0/2\}$, ${\bf k}_3=\{-k_0/2,-
\sqrt{3}k_0/2\}$,
and $A_j$ and $B_j$ ($j=1,2,3$) are slow functions
of the variables $x_j$ and $y_j$ that form orthogonal coordinate
systems with abscissa axes directed along the
wave vectors ${\bf k}_j$. Such a representation for the functions
$\eta$ and $\psi$ corresponds to a hexagonal structure of the perturbed
surface.

Using these relationships for the perturbations
$\eta$ and $\psi$
we can approximate the integral operator
$\hat{k}$ that
enters into the Hamiltonian (12) by a differential operator.
Let us use the following property:
$$
\hat{k}e^{i{\bf kr}}=
|{\bf k}|e^{i{\bf kr}},
$$
which is related to the fact that the Fourier transform of
the operator $\hat{k}$ is equal to the modulus of the wave vector.
Consider a plane wave of the form
$$
e^{i{\bf kr}}=e^{i(k_0x+q_x x+q_y y)},
$$
whose wave vector is close to $k_0$ (i.e.,$|q_x|\ll k_0$ and
$|q_y| \ll k_0$). The quantity $|{\bf k}|$ can be expanded in a series
in $q_x$ and $q_y$:
$$
|{\bf k}|=\sqrt{(k_0+q_x)^2+{q_y}^2}\approx
k_0+q_x+(2k_0)^{-1}{q_y}^2-
(2k_0^2)^{-1}q_x{q_y}^2-(2k_0)^{-3}{q_y}^4.
$$

This means that if we deal with a narrow (in the ${\bf k}$
space) wave packet with a carrying wave vector
${\bf k}=\{k_0,0\}$ (which can be represented in the form
$A(x,y)e^{ik_0x}$), then the operator $\hat{k}$ can be approximated
as follows:
$$
\hat{k}A(x,y,t) e^{ik_0x}
\approx
\left(k_0A-iA_x-(2k_0)^{-1}A_{yy}-i(2k_0^2)^{-1}A_{xyy}-
(2k_0)^{-3}A_{yyyy}\right)e^{ik_0x}
$$
(similar relations are obtained for the amplitudes $A_j$ in
the coordinates of $x_j$ and $y_j$).
Then, inserting the expressions for
$\eta$ and $\psi$
into the
Hamiltonian (12) and performing necessary averaging,
we find (to an accuracy of terms of a higher order of
smallness)
$$
H=\sum_{j=1}^3\int
\left(k_0|B_j|^2
- 2g\varepsilon |A_j|^2 +
\frac{g}{k_0^2}\left|\frac{\partial A_j}{\partial x_j}-
\frac{i}{2k_0}\frac{{\partial}^2 A_j}{\partial
y_j^2}\right|^2 \right)d^2r
$$
$$
-3gk_0\int(A_1A_2A_3+A_1^*A_2^*A_3^*)d^2r.
$$
The dynamic equations that describe the time evolution
of the amplitudes $A_j$ and $B_j$ are found from the relations
\cite{16}
$$
{A_j}_t=\frac{\delta  H}{\delta {B_j}^*},
\qquad
{B_j}_t=-\frac{\delta  H}{\delta {A_j}^*},
$$
where $j=1,2,3$.

By varying the expression for the averaged Hamiltonian,
we obtain the following equations for the amplitudes:
$$
{A_j}_t=k_0B_j,
$$
$$
{B_j}_t=2gk_0\varepsilon A_j+
\frac{g}{k_0}\left(\frac{\partial}{\partial x}-
\frac{i}{2k_0}\frac{{\partial}^2}{\partial y^2} \right)^2 A_j
+ 3g{k_0}^2\,\frac{A_1^*A_2^*A_3^*}{A_j^*}.
$$
By eliminating the amplitudes $B_j$ from these equations
and passing to dimensionless quantities using the
substitutions
\begin{equation}
\mbox{\bf r}\to \mbox{\bf r}/(\sqrt{2}k_0),
\quad
A_j\to A_j/k_0,\quad t\to t/\sqrt{2gk_0}, \quad
H\to 2Hg/k_0^2,
\end{equation}
we obtain the following set of equations:
\begin{equation}
{A_1}_{tt}=\varepsilon A_1 + {\hat{L}_1}^2 A_1 +
3\,A_2^*A_3^*\,/2,
\end{equation}
\begin{equation}
{A_2}_{tt}=\varepsilon A_2 + {\hat{L}_2}^2 A_2 +
3\,A_3^*A_1^*\,/2,
\end{equation}
\begin{equation}
{A_3}_{tt}=\varepsilon A_3 + {\hat{L}_3}^2 A_3 +
3\,A_1^*A_2^*\,/2,
\end{equation}
where we introduced operators
$$
\hat{L}_j=\frac{\partial}{\partial x_j}-
\frac{i}{\sqrt{2}}\frac{{\partial}^2}{\partial y_j^2},
\qquad j=1,2,3.
$$
The Hamiltonian corresponding to these amplitude
equations is written as follows:
\begin{equation}
H=\int\left[\sum_{j=1}^3\left(|{A_j}|_t^2+
|\hat{L}_j A_j|^2-
\varepsilon |A_j|^2\right)-
\frac{3}{2}(A_1A_2A_3+A_1^*A_2^*A_3^*)
\right] d^2r.
\end{equation}

Thus, we obtained equations that describe the initial
stages of the development of instability of the surface
of a conducting liquid in a near-critical field, when the
small-angle approximation is valid and the main nonlinear
interaction is the interaction of three standing
waves that form a hexagonal lattice. Note that analogous
equations describe the instability of the charged
surface of liquid helium \cite{8,17}.

\section{CRITERION FOR EXPLOSIVE INSTABILITY}

As is known, hexagonal structures on a charged surface
of various liquids are characterized by a hard
regime of excitation \cite{7,8}. For the set of equations
(14)-(16), this means the possibility of an unbounded
growth of the amplitude $A_j$ in a finite time period.
Indeed, in the simplest case, when the amplitudes are
independent of the spatial variables, are real, and are
equal to one another, i.e., $A_1 = A_2 = A_3 = A(t)$, the time
evolution of the quantity $A$ is described by an ordinary
differential equation with a quadratic nonlinearity
$$
A_{tt}=\varepsilon A+3A^2/2.
$$
Because of its influence, the amplitude grows
asymptotically ($A\to 4(t-t_c)^{-2}$),
under corresponding
initial conditions, i.e., the magnitude of $A$ becomes infinite
at the moment $t_c$. However, what seems to be obvious
for spatially uniform (coordinate-independent)
solutions requires to be proved in the case of arbitrary
amplitudes $A_1$ , $A_2$, and $A_3$. In particular, of a significant
interest is the situation where the initial perturbation of
the surface is localized in a certain region.

Let us show, using the method of differential inequalities,
that the nonlinear interaction of amplitude $A_j$
in terms of the model (14)-(16) results in an explosive
growth of perturbations of the surface of a conducting
liquid and find the sufficient conditions for hard excitation
of the instability. To this end, we introduce the
norm
$$
X_j(t)=\int|A_j|^2 d^2r, \qquad j=1,2,3
$$
and consider the time evolution of the following nonnegative
quantity:
$$
X=\sum_{j=1}^3 X_j.
$$

By analogy with \cite{12}, we doubly differentiate $X$ with respect to $t$
$$
X_{tt}=\sum_{j=1}^3\int\left(2|{A_j}|_t^2+{A_j}_{tt}A_j^*+
{A_j^*}_{tt}A_j\right)d^2 r
$$
$$
=\int\left[2\sum_{j=1}^3\left(|{A_j}|_t^2+
|\hat L_j A_j|^2
+\varepsilon|A_j|^2\right)+\frac{9}{2}(A_1A_2A_3+A_1^*A_2^*A_3^*)
\right]d^2 r,
$$
after substituting the corresponding right-hand sides of
the amplitude equations (14)-(16) for the multiple
derivatives ${A_j}_{tt}$ and ${A_j^*}_{tt}$.
Then, eliminating the signambiguous
cubic nonlinearity from the integrand using
the expression for the Hamiltonian (17), we obtain the
following relation:
\begin{equation}
X_{tt}+3H=-\varepsilon X+
5\sum_{j=1}^3\int\left[{|A_j|}_t^2+
|\hat L_j A_j|^2 \right] d^2r.
\end{equation}
Now, our problem is to approximate the right-hand
side of Eq.~(18) using the magnitude of $X$ and thereby
obtain an ordinary differential inequality.
From the
known Cauchy-Bunyakowsky integral inequality for
the functions
$|A_j|$ and $|{A_j}|_t$,
$$
\left[\int{|A_j|}^2 d^2r\right]\cdot\left[\int{|A_j|}_t^2
d^2r\right] \geq\left[\int{|A_j|\cdot|A_j|}_t d^2r\right]^2,
$$
it follows that $\int{|A_j|}_t^2 d^2r\geq {X_j}_t^2/(4X_j)$.
Taking also into
account the obvious relations $\int |\hat L_j A_j|^2 d^2r\geq 0$ for
$j=1,2,3$, we obtain from Eq.~(18)
\begin{equation}
X_{tt}+3H\geq-\varepsilon X+
\frac{5}{4}\sum_{j=1}^3\frac{{X_j}_t^2}{X_j}.
\end{equation}

Then, note that, as a consequence of the algebraic
Cauchy inequality, the following relation is valid:
$$
\left[\sum_{j=1}^3 X_j\right]\cdot
\left[\sum_{j=1}^3{{X_j}_t}^2/X_j\right]\geq
\left[\sum_{j=1}^3 {X_j}_t\right]^2,
$$
and, correspondingly, we have
$$
\sum_{j=1}^3{X_j}_t^2/X_j\geq{X_t}^2/X.
$$
Substituting the latter inequality into (19), we obtain
an ordinary differential inequality
\begin{equation}
X_{tt}+3H\geq-\varepsilon X+\frac{5}{4}\frac{{X_t}^2}{X},
\end{equation}
which will be the object of our consideration below.
Note that analogous inequalities arise when deriving
sufficient collapse criteria for various nonlinear partial
differential equations \cite{10,11,12,13,14}.

The introduction of a new function $Y=X^{-1/4}$ permits
us to rewrite the inequality (20) in the form of Newton's
second law
\begin{equation}
Y_{tt}\leq-\frac{\partial P(Y)}{\partial Y}, \quad P(Y)=-
\frac{1}{8}\left(\varepsilon Y^2+H Y^6\right),
\end{equation}
where $Y$
plays the role of the coordinate of a "particle"
and $P$ is its potential energy.

Let the velocity of the "particle" $Y_t$ be negative (in
this case $X_t>0$). Then, multiplying (21) by $Y_t$, we
obtain
$$
U_t(t)\geq 0, \qquad U(t)={Y_t}^2/2+P(Y),
$$
i.e., the "particle" gains an energy $U$ upon motion. It is
understandable that the sufficient criterion for $Y$ to
become zero and, correspondingly, for $X$ to become
infinity is the condition that the "particle" encounters
no potential barrier even if $U_t=0$,
which corresponds
to the equality sign in (2). The explosive growth of
amplitudes takes place under the following conditions:
\begin{itemize}
\item[(a)]
at $\varepsilon<0$ and $H>0$ if
$Y(t_0)<|\varepsilon|^{\frac{1}{4}}/(3)^{\frac{1}{4}}$ and
$12U(t_0)\leq|\varepsilon|^{\frac{3}{2}}/(3H)^{\frac{1}{2}}$;
\item[(b)]
at $\varepsilon<0$ and $H>0$ if
$12U(t_0)>|\varepsilon|^{\frac{3}{2}}/(3H)^{\frac{1}{2}}$;
\item[(c)]
at $\varepsilon<0$ and $H\leq 0$;
\item[(d)]
at $\varepsilon\geq 0$ if $U(t_0)>0$,
\end{itemize}
In (a)-(d), $t=t_0$ corresponds to the starting time
moment. In this case, the moment $t_c$,
at which the perturbation
amplitudes become infinite is estimated as
follows:
$$
t_c\leq t_0+\!\!\int\limits_0^{Y(t_0)}\!\!\frac{dY}
{\sqrt{2U(t_0)-2P(Y)}}.
$$
Note that the condition $Y_t(t_0)<0$ in cases (a) and (c)
is by no means necessary: after the reflection from a
potential wall, the "particle" reaches the point $Y=0$.
The above conditions (a)-(d) can be considered as sufficient
criteria of the instability of the surface of a conducting
liquid with respect to perturbations of a finite
amplitude, which distinguishes it from the simplest criterion
of linear instability $E > E_c$, which was derived
based on the assumption that the perturbations are infinitely
small. Note also that conditions (a)-(c) refer to
the case of subcritical external fields ($E > E_c$), when the
flat surface of the conducting liquid is stable in the linear
approximation, i.e., we deal with a hard excitation
of an electrohydrodynamic instability.

Thus, if conditions (a)-(d) are fulfilled, Eqs.~(14)-(16)
describe an infinite growth of amplitudes $A_j$. In this
case, the applicability of the model (14)-(16) to the
description of the development of an electrohydrodynamic
instability is restricted by the condition of the
smallness of the amplitudes: in the order of magnitude,
the absolute values of the amplitudes
$|A_j|$ should not
exceed the magnitude of the parameter of supercriticality
$\varepsilon$.
Otherwise, the model cannot be restricted to the
consideration of only three-wave processes. As to the
higher-order wave processes, there arises a question of
whether they will lead to a stabilization of the instability
or will favor an explosive growth of perturbations
(the experimental data of \cite{18} and the results of numerical
calculations \cite{3,4} evidence in favor of the latter situation).
The complexity of the estimation of their influence
is related to the fact that the contribution of higherorder
nonlinearities becomes comparable with the contribution
of quadratic nonlinearities of the model
(14)--(16),
only if the amplitude of surface perturbations is
close to the characteristic length of the wave. But in this
case conditions of the applicability of our approach to
the description of the near-critical dynamics of the
charged surface of liquid metal based on the construction
of amplitude equations become violated.
Nevertheless,
it is possible to reveal the influence of the
higher-order nonlinearities by considering one-dimensional
and square lattices of surface distributions, for
which the three-wave interactions degenerate and the
dominating interactions are the four-wave ones.

\section{FOUR-WAVE INTERACTIONS}

Let us consider perturbations of the boundary of a
conducting liquid with such symmetries for which the
effect of three-wave processes is negligible. Thereby,
we in pure form separate the four-wave interactions that
determine the character of the electrohydrodynamic
instability at its advanced stages.

First of all, we consider the near-critical behavior of
the charged surface of a liquid metal in the assumption
of a quasi-one-dimensional character of the arising
wave. Let the wave vector be parallel to the abscissa
axis. We pass to envelopes by using substitutions
$$
\eta(x,y,t)=
A(x,y,t)e^{ik_0x}+
A_0(x,y,t)e^{2ik_0x}+\mbox{c.c.},
$$
$$
\psi(x,y,t)=
B(x,y,t)e^{ik_0x}+
B_0(x,y,t)e^{2ik_0x}+\mbox{c.c.},
$$
in which the  $k_0\leftrightarrow 2k_0$ interaction is taken into
account. Here, $A$, $B$, $A_0$ and $B_0$ are slowly varying functions
of the spatial variables $x$ and $y$.
Substituting these
relations into the Hamiltonian (12), we find (to an accuracy
of terms of the fourth order of smallness)
$$
H=H^{(2)}+H^{(3)}+H^{(4)},
$$
$$
H^{(2)}=\int\left[k_0|B|^2-2g\varepsilon|A|^2+
\frac{g}{k_0^2}\left|\frac{\partial A}{\partial x}-
\frac{i}{2k_0}\frac{{\partial}^2 A}{\partial y^2}\right|^2
+g|A_0|^2+2k_0|B_0|^2\right]d^2r,
$$
$$
H^{(3)}=-2gk_0\int\left[A^2{A_0}^*+{A^*}^2A_0\right] d^2r,
$$
$$
H^{(4)}=\frac{5}{4}\,gk_0^2\int |A|^4 d^2r.
$$
The amplitude equations for the perturbations of the
free surface are written in the Hamiltonian form as
$$
\frac{\partial A}{\partial t}=\frac{\delta H}{\delta B^*},
\qquad
\frac{\partial B}{\partial t}=-\frac{\delta  H}{\delta A^*},
\qquad
\frac{\partial A_0}{\partial t}=\frac{\delta  H}{\delta
{B_0}^*},
\qquad
\frac{\partial B_0}{\partial t}=-\frac{\delta  H}{\delta
{A_0}^*}.
$$
By varying the function $H$, and then eliminating the
quantities $B$ and $B_0$, we obtain equations of the form
\begin{equation}
A_{tt}=2gk_0\varepsilon A+\frac{g}{k_0}
\left(\frac{\partial}{\partial x}-
\frac{i}{2k_0}\frac{{\partial}^2}{\partial y^2}\right)^2 A +
4g{k_0}^2 A^* A_0 - \frac{5}{2}\,g{k_0}^3 A^2 A^*,
\end{equation}
\begin{equation}
{A_0}_{tt}=-2gk_0A_0+4g{k_0}^2A^2.
\end{equation}
Since the characteristic times of changes in the
amplitudes at small supercriticalities are small
($\omega^2\sim\varepsilon$),
we neglect the derivatives with respect to time in
Eq.~(23). Then, the quantity $A_0$ can be expressed
through the amplitude $A$ that plays the role of an order
parameter
$$
A_0\approx 2k_0 A^2.
$$
Using this relation, we eliminate $A_0$
from Eq.~(22)
and pass to dimensionless quantities using scalings (13)
to obtain for the complex amplitude $A$:
\begin{equation}
A_{tt}=\varepsilon A+{\hat{L}_1}^2A+sA\,|A|^2,
\qquad s=11/4,
\end{equation}
to which the following expression for the Hamiltonian corresponds:
\begin{equation}
H=\int \left[|A_t|^2+\left|{\hat{L}_1}A\right|^2-\varepsilon
|A|^2 - s|A|^4/2\right] d^2r.
\end{equation}

Note that, when neglecting the dependence of the
amplitude $A$ on $y$, Eq.~(24) becomes a nonlinear Klein-Gordon
equation, i.e., corresponds to the so-called
$|\psi|^4$ model. In this form, it can be obtained from the equation
for one-dimensional perturbations of the charged
surface of liquid helium \cite{8} in the limit of the complete
shielding of the field under the surface. Note also that if
we neglect transverse modulations, then Eq.~(24) coincides
with that obtained in the Kelvin-Helmholtz theory
of instability for the case of a small ratio of the densities
of the top and bottom liquids \cite{12}. This is due to
the identity of the mathematical description of the planar
potential flow of an incompressible liquid and a
two-dimensional distribution of an electric field in the
absence of spatial electric charges. The allowance for
higher-order terms in the expansions in surface perturbations
violates this analogy.

Since the term $|A_t|^2$ that is responsible for the kinetic
energy and the term $|A|^4$ that is responsible for the fourwave
processes enter into the integrand of the Hamiltonian
(25) with the opposite signs, Eq.~(24) admits infinite solutions.
This means that the cubic nonlinearity in
Eq.~(24) does not stabilize the linear instability but, on
the contrary, enhances it, leading, under certain conditions,
to an explosive growth of the amplitude $A$ of perturbations
of the conducting-liquid boundary.

Another possible case when the three-wave processes
are degenerate is the interaction of two standing
waves whose wave vectors
${\bf k}_1$ and ${\bf k}_2$  are turned with
respect to one another by an angle $\pi/2$
(the vector's
coordinates are  ${\bf k}_1=\{k_0,0\}$ and ${\bf k}_2=\{0,k_0$\}).
Let us represent
the perturbation of the surface
$\eta$ in the form
$$
\eta({\bf r},t)= \sum_{j=1}^2\left[
A_je^{i{\bf k}_j{\bf r}}+2k_0 A_j^2e^{2i{\bf k}_j{\bf
r}}\right]
$$
$$
+\left(12\sqrt{2}+16\right)k_0
\left[A_1A_2 e^{i({\bf k}_1+{\bf k}_2){\bf r}}+
A_1A_2^* e^{i({\bf k}_1-{\bf k}_2){\bf
r}}\right]+\mbox{c.c.},
$$
and the perturbation of the velocity potential at the liquid
boundary $\psi$ in the form
$$
\psi({\bf r},t)= \sum_{j=1}^2\left[k_0^{-1}\!{A_j}_t
e^{i{\bf k}_j{\bf r}}+2A_j {A_j}_t e^{2i{\bf k}_j{\bf
r}}\right]
$$
$$
+\left(6+4\sqrt{2}\right)\left[\left(A_1A_2\right)_t
e^{i({\bf k}_1+{\bf k}_2){\bf r}}+
\left(A_1{A_2}^*\right)_t
e^{i({\bf k}_1-{\bf k}_2){\bf r}}\right]+\mbox{c.c.},
$$
where we took into account the nonlinear interactions
of the fundamental spatial harmonic $k_0$ with mixed harmonics $2k_0$
and $\sqrt{2}k_0$.
This representation for the functions $\eta$ and $\psi$
corresponds to the symmetry of a square
lattice.

Proceeding by analogy to the above-considered
quasi-one-dimensional case, we obtain, after passing to
dimensionless quantities, the following dynamic equations:
$$
{A_1}_{tt}=\varepsilon A_1 + {\hat{L}_1^2} A_1 +
s A_1|A_1|^2 + \sigma A_1|A_2|^2,
$$
$$
{A_2}_{tt}=\varepsilon A_2 +
{\hat{L}_2}^2 A_2 + s A_2|A_2|^2 +
\sigma A_2|A_1|^2,
$$
where $\sigma=32\sqrt{2}+65/2$, and the following designations
are introduced:
$$
\hat{L}_1=\frac{\partial}{\partial x}-
\frac{i}{\sqrt{2}}\frac{{\partial}^2}{\partial y^2}
\quad\mbox{and}\quad
\hat{L}_2=\frac{\partial}{\partial y}-
\frac{i}{\sqrt{2}}\frac{{\partial}^2}{\partial x^2}.
$$

The integral of motion for these equations, corresponding
to the conservation of the total energy of a
conservative system, is given by the expression
$$
H=\sum_{j=1}^2\int\left(|{A_j}|_t^2+
|\hat{L}_j A_j|^2-
\varepsilon |A_j|^2-s|A_j|^4/2\right)d^2 r-
\sigma\int|A_1|^2|A_2|^2 d^2 r.
$$
The first term on the right-hand side of this functional
coincides in its structure with the Hamiltonian
(25) for the quasi-one-dimensional wave. The last term
is responsible for the nonlinear interaction of a pair of
the waves studied. Note that the coefficient
$\sigma$ before
this term exceeds the coefficient $s$ by more than an
order of magnitude. This means that the contribution of
the interaction ${\bf k}_1\leftrightarrow{\bf k}_2$
is determining and, consequently,
the square structure of the surface perturbations
is much more favorable than the one-dimensional
one.

In any case, for both the square and one-dimensional
lattice (the latter can be considered as a partial
case of the square lattice, corresponding to the condition
$A_2=0$), the four-wave interactions will favor the
development of an instability rather than suppress it.
Conditions for an explosive growth of the amplitudes
$A_1$ and $A_2$ can be obtained by considering the evolution
of the norm
$$
X=\int \left(|A_1|^2+|A_2|^2\right)d^2r.
$$
Acting by analogy with Section 5, we obtain the
majorizing inequality
$$
X_{tt}+4H\geq-2\varepsilon X+\frac{3}{2}\frac{X_t^2}{X},
$$
which coincides with that considered in \cite{12}. The introduction
of the variable $Y=X^{-1/2}$ reduces the problem to
the analysis of the motion of a "particle" with a coordinate
$Y$ in a potential well $P(Y)$
$$
Y_{tt}\leq-\frac{\partial P(Y)}{\partial Y}, \qquad P(Y)=-
\frac{1}{2}\left(\varepsilon Y^2+HY^4\right).
$$
Analyzing this inequality for the case where the
velocity of the "particle" at the initial time moment
$t=t_0$ is directed toward the origin (i.e.,
$Y_t(t_0)<0$), it can
easily be revealed that the quantity $Y$ vanishes, first, at
$\varepsilon>0$, if $U(t_0)>0$, second, at $\varepsilon<0$ and $H<0$, and
third, at $\varepsilon<0$ and $H>0$, if
$U(t_0)>\varepsilon^2/(8H)$ or
$Y^2(t_0)<|\varepsilon|/(2)$. Here, $U(Y)$, just as in Section 5,
denotes the total
mechanical energy of the "particle." Under the above
conditions, the norm $X$ become infinite in a finite time,
which just corresponds to an explosive growth of the
amplitudes in the result of four-wave interactions.

All this suggests that the higher-order nonlinearities
will not suppress the explosive growth of amplitudes in
the model (14)-(16). But in this case the above integral
criteria (a)-(d) may be considered as sufficient criteria
of the explosive growth of perturbations of the surface
of a liquid metal in an external electric field.

\section{CONCLUSION}

The main result of this work is the construction of
sufficient integral criteria of instability for the free surface
of an ideally conducting liquid in a near-critical
external electric field. These criteria represent a generalization
of the known condition of linear instability
($E > E_c$) to the case where the amplitudes of perturbations
of the field of velocities and of the shape of the
surface are finite. The criteria found are dynamic in the
sense that they take into account the effect of the velocity
distribution in the medium at the initial time
moment; the role of the stored kinetic energy can be
decisive in the case of the hard mechanism of instability.

An analysis of three-wave and four-wave nonlinear
interactions (this corresponds to the allowance for quadratic
and cubic nonlinearities in the amplitude equations)
showed that the development of the electrohydrodynamic
instability has an explosive character, i.e.,
leads to the appearance of singularities in the solutions
in a finite time. This conclusion qualitatively agrees
with the results of numerical simulation of the development
of the instability of the boundary of liquid metal:
it was shown in \cite{4} that the curvature of the surface
increases according to a power law characteristic of the
explosive instability and causes the formation of specific
features of a cusplike type.

Note in conclusion that
the criteria of hard instability analogous to the criteria
(a)-(d) can also be obtained for insulating liquids with
induced surface charges \cite{9}, for insulating liquids with
free surface charges (liquid helium and liquid hydrogen
in an electric field refer to this category) \cite{9, 19}, and for
ferromagnetic liquids in a vertical magnetic field.

\bigskip
This work was supported in part by the Russian
Foundation for Basic Research (project no. 00-02-
17428) and by the INTAS (project no. 99-1068). We are
grateful to N.B. Volkov and A.M. Iskol'dskii
for interest in our work; N.M. Zubarev is also grateful to
E.A. Kuznetsov for stimulating discussions.

\end{document}